\documentclass[11pt]{article}
\usepackage{hyperref}
\usepackage{amssymb}
\usepackage{graphicx}
\usepackage{subcaption}
\usepackage{float}
\usepackage{slashed}
\usepackage{amsmath}
\usepackage{amssymb}
\usepackage{tikz}
\usepackage{fixmath}
\usepackage{tabularx}
\usepackage{breqn}
\usepackage{authblk}
\usepackage{array}
\usepackage{dsfont}
\usepackage{cite}
\usepackage{physics}
\usepackage[section]{placeins}
\usepackage[a4paper, total={6.8in, 10in}]{geometry}
\numberwithin{equation}{section}
\newcommand{\be}{\begin{equation}}
\newcommand{\ee}{\end{equation}}
\newcommand{\bea}{\begin{eqnarray}}
\newcommand{\eea}{\end{eqnarray}}
\newcommand{\ba}{\begin{aligned}}
\newcommand{\ea}{\end{aligned}}
\begin{document}
\title{Geodesic motion of particles in the vicinity of the $\kappa$-deformed Schwarzchild Black Hole}

\author[1]{Dilip Kumar \thanks{21phph03@uohyd.ac.in}}
\author[1]{Suman Kumar Panja \thanks{sumanpanja19@gmail.com, 19phph17@uohyd.ac.in}}
\author[2,3]{Abhisek Saha\thanks{saha@pku.edu.cn}}  
 \author[1]{Soma Sanyal \thanks{somasanyal@uohyd.ac.in} }
\affil[1]{School of Physics, University of Hyderabad, Central University P.O, Hyderabad-500046, Telangana, India}
\affil[2]{Center for High Energy Physics, Peking University, Beijing 100871, China}
\affil[3]{School of Physics, Peking University, Beijing 100871, China}
\date{}
\maketitle

\begin{abstract}

In this study, we investigate the geodesic motion of a test particle around the Schwarzchild black hole in a $\kappa$-deformed space-time. We compute a modified Lagrangian to obtain the $\kappa$-deformed effective potential and find the particle trajectories based on the constants of motion. For the same value of angular momentum, we obtain a significant deformation in the orbits of the particles due to the non-commutativity of the $\kappa$-deformed space-time. The deformation parameter becomes more significant for higher values of the angular momentum. The radius of the individual trajectories become smaller and their velocities decrease compared to the commutative case. The radius of the innermost stable circular orbit ($r_{ISCO}$) is also found using the modified effective potential. Though the equations get modified due to the non-commutativity of the $\kappa$-deformed space-time, the $r_{ISCO}$ remains the same. We then study a large number of freely streaming particles moving in this $\kappa$-deformed space-time and analyze the movement of these particles around the black hole due to the non-commutativity of the space-time. We concentrate on particles with different angular momentum moving around the black hole. We find that the motion of the particles are modified due to the non-commutativity of the space-time. The particles move slower along their respective trajectories in the deformed space-time. So, they remain closer to the black hole for a longer period of time, indicating that the accretion of freely streaming particles around the black hole would be modified by the non-commutavity of the space-time.

\end{abstract}

\section{Introduction} 
 A star is a massive self-luminous astronomical body of gas that shines by radiation derived from its internal energy sources. For a stable star, the inward force of gravity and the outward pressure of the hot gas are in balance. The attractive gravitational pull towards a stable star's massive core is intense. This is countered by the outward pressure created by nuclear fusion. When this fusion stops, the outer pressure drops and the colossal star collapses, resulting in a supernova explosion. For a star with mass greater than 20 solar mass, the core collapses to a singularity 
creating a black hole. In the case of slightly less massive stars, the collapse results in the formation of a neutron star \cite{cameron}. The collapse of stars with less than four solar masses leads to the formation of a white dwarf \cite{chandra}. These diverse outcomes showcase the remarkable variety in the aftermath of supernova explosions, contributing to our understanding of the lifecycle of stars. 

The gravitational force near these kind of superdense compact objects is very high. Black holes represent one of the most extreme environments in the universe, where gravitational forces are extremely strong. The geodesic motion of a particle around a black hole offers a unique testing ground for the predictions of
Einstein's general relativity. Observations of the motion of objects near black holes can confirm or
refine our understanding of gravity in extreme conditions and verify the accuracy of general
relativity's predictions. Various studies \cite{Mit, Page, Cruz, Abdu,perez,olvera,mandal} have been taken in this direction. Accretion of simple polytropic gas onto a steady-state spherically symmetric Schwarzschild black hole was studied in ref.\cite{Mit}. In ref. \cite{Page}, the geodesic study of test particles has been done. It is shown that geodesic motion is completely integrable for rotating black holes. In \cite{Cruz}, authors studied the geodesic structure of the Ads Black holes using the effective potential. Investigation of the motion of charged particles around the rotating black hole in a Braneworld is done in \cite{Abdu}. The authors have explicitly calculated the effective potential and radius for the innermost stable circular orbit (ISCO). The dynamics of the test particles around rotating black holes have been studied in \cite{olvera}. This study described the properties of the geodesic using the Hamilton-Jacobi approach for a Heyward-rotating black hole. In ref.\cite{mandal},the geodesic motion near an improved Schwarzchild black hole is investigated. In it, the authors have discussed the stability of circular orbits using the energy and angular momentum of massive particles. The authors also obtained the ISCO of the massive particles. In \cite{gat,biswas}, motions of the test particles were analyzed in different space-times. Geodesic study in conical Schwarzchild space-time and conical Lense-Thirring space-time is discussed in \cite{gat}. In \cite{biswas}, the authors studied the motion of the test particles around the cosmic string (connected to a black hole) using geodesic equations and the Hamilton-Jacobi formalism. The geodesic motion of test objects has been studied in the presence of the cosmological constant \cite{cal} too. This article studied the presence of the cosmological constant on the geodesic motion of the particles around the Reissner-Nordstrom-de-Sitter and Kerr-de-Sitter black holes. All these studies have significantly improved our understanding of the motion of particles around compact and massive objects in the universe.

Quantum gravity models are expected to modify the space-time structure when the gravitational field is very strong. Thus, it is worth emphasizing that not only in the large length scale but even in the Planck scale, the phenomenon of geodesic motion may lead to provide important insights in testing the physical principles. Many studies have therefore been initiated to explore the quantum gravity effects in the geodesic motion\cite{yang,sunandan,kappa-geod} of particles. In \cite{yang}, quantum gravity effects have been studied in the process of the accretion of polytropic matter onto a Schwarzschild black hole. Accretion of matter onto Schwarchild black hole in non-commutative space-time has been investigated in ref. \cite{sunandan}. In this study, the authors analyzed the matter creation in the Moyal space-time and found that it increases rapidly with the strength of the non-commutative parameter. In \cite{kappa-geod}, corrections to the geodesic equation are obtained due to the non-commutativity of the $\kappa$-deformed space-time. The non-commutative effect is interpreted as an extra drag that acts during the motion of the particle in
this $\kappa$-deformed curved space-time. 

One of the oldest and most intriguing topics in physics is the development of the quantum gravity model and the understanding of the nature of gravity at the microscopic scale. Several approaches have been taken to investigate gravity at the microscopic scale, such as String theory, loop quantum gravity, causal sets, dynamical triangulation, and non-commutative geometry \cite{connes, douglas, rov, sorkin, Glikman, dop, am,madore,seiberg-witten}. The existence of a fundamental length scale introduced by these approaches is a characteristic feature of quantum gravity \cite{Glikman, dop}. Non-commutative geometry inherently incorporates this length scale and thus provides a testing environment to build quantum gravity models. Thus, it is of paramount interest to study how non-commutative space-time will affect the models built in the framework of General-relativity and Cosmology. 

In recent times various models on non-commutative space-times and their implications have been studied extensively \cite{douglas, dop,wess1,chaichian,kappa1,dimitrijevic,dasz,mel1, carlson,amorim}. One well-studied non-commutative space-time is Moyal space-time \cite{douglas}. Moyal space-time violates the Lorentz symmetry, and its symmetry algebra has been realized using Hopf algebra \cite{wess1,chaichian}. Another non-commutative space-time is $\kappa$-deformed space-time, whose coordinates obey Lie algebraic type space-time commutation relation. The corresponding symmetry algebra has been defined using the Hopf algebra \cite{kappa1,dimitrijevic,dasz,mel1}. In $\kappa$-space-time, the coordinates satisfy the following commutation relations
\begin{equation}
[\hat{x}^i,\hat{x}^j]=0,~~~[\hat{x}^0, \hat{x}^i]=ia\hat{x}^i,~~~a=\frac{1}{\kappa}. \label{com2}
\end{equation}
where $a$ is the deformation parameter having dimension of $\textit{length}$. Since quantum theories of non-commutative space-time inherently incorporate fundamental length scale, studying physical phenomena in the presence of the fundamental length scale is important. 

Various studies in non-commutative gravity and corresponding physics have been analyzed. The effects of non-commutativity in cosmological objects in the $\kappa$-deformed space-time have been investigated and reported in \cite{zuhair1,zuhair2}. In \cite{zuhair1}, $\kappa$-deformed corrections to the Hawking radiation are derived using the method of Bogoliubov coefficients. Compact stars in the $\kappa$-deformed space-time have been studied using the partition function and the generalized uncertainty principle \cite{zuhair2}. Various aspects of black hole physics in non-commutative space-time were analyzed in \cite{kappa-btz,kappa-btz1}. Using the core-envelope model and $\kappa$-deformed Einstein field equations superdense stars are studied in \cite{visnhu}. The evolution of the universe using Newtonian Cosmology is studied in $\kappa$-deformed space-time \cite{suman}. The authors of \cite{neutronstar} analyzed a Neutron star model in $\kappa$-deformed non-commutative space-time and obtained the maximum mass of the neutron star. In \cite{neutronstarcosmo}, influence of the Cosmological constant on Neutron star has been investigated in $\kappa$-deformed space-time. Thus, studying various models in non-commutative space-time gives a deeper understanding of the effects of quantum gravity. The investigation of models built on the framework of general relativity and cosmology in non-commutative space-time will provide us the testing ground to study the effects of quantum gravity and leads us to understand the nature of gravity in quantum regimes. Thus, it is of intrinsic interest to investigate the impact of non-commutativity in studying the geodesic motion of an object around a black hole. In this study, we examine the motion of an object around the black hole in $\kappa$-deformed space-time.

Our aim is to investigate the behavior of a test particle near a static spherical black hole. We construct $\kappa$-deformed metric for the Scwarzchild black hole. We use this modified metric to find the geodesic equations valid up to the first order for the deformation parameter $a$. We analyze these geodesic equations to see the trajectory of a test particle around the Schwarzchild black hole in $\kappa$-deformed space-time. We examine the changes in the particle's motion due to the presence of the deformation parameter. Next, using the $\kappa$-deformed metric, we obtain the modified Lagrangian, and using this, we derive the $\kappa$-deformed effective potential valid up to the first order in $a$. We get different trajectories depending on the values of constants of motion. We are particularly interested in the variation of the trajectories with the angular momentum. At higher angular momentum we get significant change in the orbits of the particles. Using a modified form of the effective potential, we find the radius of the innermost stable circular orbit ($r_{ISCO}$).

Once we have studied the trajectories of the individual particles, we use a multi-particle code to generate a large number of particle having different initial positions, angular momentum and velocity. The large number of particles follow the geodesic equations and are equivalent to freely streaming particles in the space-time of the black hole. The particles move along the geodesics in the undeformed and the deformed space-time. We find that in the undeformed space-time the particles with higher angular momentum move away from the black hole due to larger radius of the scattering orbits. For the deformed space-time, however the orbits are more compact and the streaming particles remain closer to the black holes.      
 
The organization of this paper is as follows. The next section briefly reviews $\kappa$-deformed space-time and its symmetry algebra. In section 3, we choose a specific realization for $\kappa$-deformed space-time. This helps us write the non-commutative variables in terms of the commutative variables and their derivatives and the deformation parameter $a$. Using this in the generalized commutation relation for the phase space coordinates, we obtain the $\kappa$-deformed metric for the Schwarzchild black hole. In section 4, we study the deformed geodesic motion of an object and particles's distribution due to the non-commutativity of the $\kappa$-deformed space-time. We construct $\kappa$-deformed geodesic equations in subsection 4.1. Using these equations of motion, we investigate the motion of an object around the Schwarzchild black hole in the $\kappa$-deformed space-time. Further, we find the $\kappa$-deformed effective potential and the radius of the innermost stable circular orbit ($r_{ISCO}$) of an object in subsection 4.2. In subsection 4.3, we study the distribution of the freely streaming particles near the $\kappa$-deformed Schwarzchild black hole. 
In section 5, we summarize and conclude the paper.

\section{Kappa Deformed Space-Time}

In this section, we summarize the details of the $\kappa$-deformed space-time. The $\kappa$-deformed space-time is a Lie-algebraic type non-commutative space-time, whose coordinates satisfy Eq.(\ref{com2}). The $\kappa$-deformed coordinate $\hat{x}_{\mu}$ is written in terms of the commutative coordinate $x_{\mu}$ and its derivatives $\partial_{\mu}$ as  \cite{mel1}
\begin{equation}\label{ksp-2}
\begin{split}
 \hat{x}_0=&x_0\psi(A)+iax_j\partial_j\gamma(A)\\
 \hat{x}_i=&x_i\varphi(A),
\end{split}
\end{equation}
where $A=ia\partial_0=ap^{0}$, and $\psi$, $\gamma$, and $\varphi$ are functions of $A$, satisfy the following conditions
\begin{equation}\label{ksp-3}
 \psi(0)=1,~\varphi(0)=1.
\end{equation}   
We substitute Eq.(\ref{ksp-2}) in Eq.(\ref{com2}) and we obtain
\begin{equation}\label{ksp-4}
 \frac{\varphi'(A)}{\varphi(A)}\psi(A)=\gamma(A)-1,
\end{equation}
with $\varphi^{\prime}=\frac{d\varphi}{dA}$. Two possible realizations of $\psi(A)$ are $\psi(A)=1$ and $\psi(A)=1+2A$  \cite{mel1}. In this work, we choose $\psi(A)=1$. Thus, Equations (\ref{ksp-2}) and  (\ref{ksp-4}) become
\begin{equation}\label{ksp-5}
\begin{split}
 \hat{x}_0=&x_0+iax_j\partial_j\gamma(A)\\
 \hat{x}_i=&x_i\varphi(A),
\end{split}
\end{equation}
and
\begin{equation}\label{ksp-6}
 \frac{\varphi'(A)}{\varphi(A)}=\gamma(A)-1.
\end{equation} 
Here, the allowed choices of $\varphi$ are $e^{-A}, e^{-\frac{A}{2}}, 1, \frac{A}{e^A-1}$, etc.  \cite{mel1}. For our study, we choose $\varphi(A)=e^{-A}$.

\section{Kappa-deformed Schwarzchild metric}

In this section, we construct the $\kappa$-deformed metric for any space-time. We use the generalized commutation relation between the $\kappa$-deformed coordinates and the corresponding conjugate momentum\cite{kappa-geod}. Next, we find the $\kappa$-deformed metric for the isotropic Schwarzchild black hole. We start with the generalized commutation relation for the $\kappa$-deformed phase space coordinates \cite{kappa-geod} as
\begin{equation}\label{N1}
 [\hat{x}_{\mu},\hat{P}_{\nu}]=i\hat{g}_{\mu\nu}, 
\end{equation} 
where $\hat{g}_{\mu\nu}(\hat{x}^{\alpha})$ is the deformed metric in $\kappa$-space-time. The $\kappa$-deformed phase-space coordinates are realized as \cite{kappa-geod},
\begin{equation}\label{N2}
 \hat{x}_{\mu}=x_{\alpha}\varphi^{\alpha}_{\mu}, \,\hat{P}_{\mu}=g_{\alpha\beta}(\hat{y})p^{\beta}\varphi^{\alpha}_{\mu},
\end{equation}
where $\hat{P}_{\mu}$ is the $\kappa$-deformed generalised momenta corresponding to the  $\hat{x}_{\mu}$, the non-commutative co-ordinates. $p_{\mu}$ is the canonical conjugate momenta corresponding to the commutative coordinate $x_{\mu}$. Note here that in the limit, $a \to 0$, we find $\hat{x}_{\mu}\to x_{\mu}$ and $\hat{P}_{\mu}\to p_{\mu}$. Note that in Eq.(\ref{N2}), we have introduced another set of $\kappa$-deformed space-time coordinates $\hat{y}_{\mu}$. This $\hat{y}_{\mu}$ is assumed to commute with $\hat{x}_{\mu}$, i.e., $[\hat{y}_{\mu},\hat{x}_{\nu}]=0$. 

Substituting Eq.(\ref{N2}) in Eq.(\ref{com2}) we find a particular realisation for $\varphi_{\mu}^{\alpha}$ as
\begin{equation}\label{N3}
 \varphi _0^0=1, \, \varphi _i^0=0, \, \varphi_0^i=0, \, \varphi _j^i=\delta _j^i e^{-ap^0}. 
\end{equation}
We have assumed that the coordinates $\hat{y}_{\mu}$ will satisfy the $\kappa$-deformed space-time commutation relation (given in Eq.(\ref{com2})) as 
\begin{equation}
[\hat{y}_0,\hat{y}_i]=ia\hat{y}_i,~~~[\hat{y}_i,\hat{y}_j]=0. \label{N3b}
\end{equation}
Now $\hat{y}_{\mu}$ can be realized (in terms of the commutative coordinates and corresponding conjugate momenta) as
\begin{equation}\label{N4}
 \hat{y}_{\mu}=x_{\alpha}\phi_{\mu}^{\alpha}.
 \end{equation}
Using Eq.(\ref{N3b}) and $[\hat{x}_{\mu},\hat{y}_{\nu}]=0$, one find $\phi_{\mu}^{\alpha}$ as (see \cite{zuhair1,kappa-geod} for details)
\begin{equation}\label{N5}
 \phi_{0}^{0}=1,~\phi_{i}^{0}=-ap^i,~\phi_{0}^{i}=0,~\phi_{i}^{j}=\delta_{i}^{j}.
\end{equation}
Thus using Eq.(\ref{N5}) in Eq.(\ref{N4}) we find the explicit form of $\hat{y}_{\mu}$ to be
\begin{equation}\label{N6}
 \hat{y}_0=x_0-ax_jp^j,~~
\hat{y}_i=x_i.
 \end{equation}
Using the above equation in Eq.(\ref{N2}) and substituting $\hat{x}_{\mu}$ and $\hat{P}_{\mu}$ in Eq.(\ref{N1}), one find the $\kappa$-deformed metric as \cite{zuhair1}
\begin{equation}\label{N7}
 [\hat{x}_{\mu},\hat{P}_{\nu}] \equiv i\hat{g}_{\mu\nu}=ig_{\alpha\beta}(\hat{y})\Big(p^{\beta}\frac{\partial \varphi^{\alpha}_{\nu}}{\partial p^{\sigma}}\varphi_{\mu}^{\sigma}+\varphi_{\mu}^{\alpha}\varphi_{\nu}^{\beta}\Big). \end{equation}
Note that $g_{\mu\nu}(\hat{y})$ in the above is obtained by replacing commutative coordinates with non-commutative coordinates $\hat{y}_{\mu}$ in the expression of commutative metric. This $\hat{y}_{\mu}$ can be expressed in terms of commutative coordinates and corresponding conjugate momenta using Eq.(\ref{N6}).

We substitute Eq.(\ref{N3}) in Eq.(\ref{N7}) and find the components of $\hat{g}_{\mu\nu}$ as
\begin{equation}\label{N9}
\begin{aligned}
\hat{g}_{00}&=g_{00}(\hat{y}),\\
\hat{g}_{0i}&=g_{0i}(\hat{y}) e^{-2ap^{0}}-ag_{im}(\hat{y})p^m e^{-ap^{0}},\\ 
\hat{g}_{i0}&=g_{i0}(\hat{y})e^{-ap^{0}},\\
\hat{g}_{ij}&=g_{ij}(\hat{y})e^{-2ap^{0}}.
\end{aligned}
\end{equation}
The explicit form of the line element in $\kappa$-deformed space-time is defined \cite{zuhair1} as 
\begin{equation}\label{N11}
\begin{aligned}
d\hat{s}^2&=g_{00}(\hat{y})dx^0dx^0+\Big(g_{0i}(\hat{y})\big(1-ap^0\big)-ag_{im}(\hat{y})p^m\Big)e^{-2ap^{0}}dx^0dx^i\\&+g_{i0}(\hat{y})e^{-2ap^{0}}dx^idx^0+g_{ij}(\hat{y})e^{-4ap^{0}}dx^idx^j.
\end{aligned}
\end{equation}
We observe that the metric components have an explicit dependency on spatial coordinates only. Thus from Eq.(\ref{N6}) and Eq.(\ref{N11}), we find $g_{\mu\nu}(\hat{y}^i)=g_{\mu\nu}(x^i)$.

In this study, we investigate the motion of a test particle around an isotropic Schwarzchild black hole. The line element corresponding to an isotropic Schwarzchild black hole is given as \cite{Buch,Diemer}
\begin{equation}
 ds^{2} = - f(r)dt^{2} + \frac{l(r)}{f(r)} \, \bigg[ dr^{2} \, + \, r^{2} \, d\theta^{2} \, + \, r^{2} \sin^{2}\theta \, d\phi^{2} \bigg]\label{N11a},
\end{equation}
where, $f(r) = \bigg( \frac{1 - \frac{M}{2 r}}{1 + \frac{M}{2 r}}\bigg)^{2}$ and $l(r)= \bigg( 1 - \frac{M^{2}}{4 r^{2}}\bigg)^{2}$ \cite{Buch,Diemer}. Since the cross terms in the isotropic Schwarzchild metric tensor given in Eq.(\ref{N11a}) are zero (i.e., $g_{0i}=0$), the $\kappa$-deformed Schwarzchild metric becomes \footnote{Under Rotations $\kappa$-deformed space-time remain invariant. For calculational simplification, we employ realization (see eq.(\ref{N6})) for non-commutative coordinate $\hat{y}_{\mu}$ in deriving the $\kappa$-deformed metric. This realization leads to non-zero off-diagonal terms in the $\kappa$-deformed metric. We set off-diagonal components in the $\kappa$-deformed metric to zero in order to retain rotational symmetry.}
\begin{equation}\label{N12}
 d\hat{s}^2=g_{00}(\hat{y})dx^0dx^0+g_{ij}(\hat{y})e^{-4ap^0}dx^idx^j. 
\end{equation}
Using Eq.(\ref{N11a}) in Eq.(\ref{N12}), we find the $\kappa$-deformed line element for isotropic Schwarzchild Black Hole as
\begin{equation} \label{N13}
 d\hat{s}^2= - f(r)dt^{2} \, + \frac{l(r)}{f(r)} \, \, e^{-4 a p^{0}} \, \bigg[ dr^{2} \, + \, r^{2} \, d\theta^{2} \, + \, r^{2} \sin^{2}\theta \, d\phi^{2} \bigg]
\end{equation}
We mention here, that in the commutative limit, i.e., $a \rightarrow 0$, the above equation reduces to the Eq.(\ref{N11a}).

\section{Geodesics in the $\kappa$-deformed space-time}
In this section, we study the geodesic motion of a test particle around the Schwarzchild black hole in the background of the $\kappa$-deformed space-time. Geodesics around the Schwarzchild black hole have previously been studied in detail. The trajectories of the particles depend on the two constants of motion the energy and the angular momentum. These have been parametrized in various ways to determine the different kinds of particle trajectories around the black hole. While ellliptical orbits have been obtained in all the cases, there is a definite apsidal precession of the orbits with time. Particles can fall into the black hole or can be scattered away from the black hole depending on their initial conditions of angular momemtum and energy. In most cases, it has been assumed that the particles have come from infinity. We now proceed to study these same geodesics assuming that the black hole is in a $\kappa$-deformed space-time.

\subsection{Test particle trajectories in  $\kappa$-deformed space-time }
In this subsection, we start with investigating the trajectory of a single test particle around a black hole in the background of the $\kappa$-deformed space-time. For this, we use the $\kappa$-deformed generalized expression of the geodesic equation as  
\begin{equation}
\frac{d^{2} \hat{x}^{\mu}}{d \tau^{2}}  +  \hat{\Gamma}^{\mu}_{\rho \sigma} \frac{d \hat{x}^{\rho}}{d \tau} \frac{d \hat{x}^{\sigma}}{d \tau}= 0.  \label{geod1}
\end{equation}

Using the $\kappa$-deformed line element for the isotropic Schwarzchild black hole given in eq.(\ref{N13}) in the above equation, we obtain the different components of the geodesic equations. 
The dependence of proper time is given by the equation 	
		\begin{equation}
		\frac{d^{2} t}{d \tau^{2}}  + e^{-a p^{0}} \frac{1}{(1 - \frac{M^{2}}{4 \, r^{2}})} \, . \, \bigg(\frac{2 \, M}{r^{2}}\bigg)  \, \frac{d \, r}{d \tau} \, . \, \frac{d \, t}{d \tau} = 0, \label{geod2}
		\end{equation}
For massive particles, we would like to plot the trajectories in the three dimensions using the spherical polar coordinates $(r, \theta, \phi)$, the equation for $r$, the radius of the orbit is obtained from  		
		\begin{multline}
		 \frac{d^{2} r}{d \tau^{2}}  + e^{5 a p^{0}}~\frac{M}{r^{2}} \Bigg(\frac{1 - \frac{M}{2 \, r}}{(1 + \frac{M}{2 \, r})^{7}} \Bigg) \bigg(\frac{d \, t}{d \tau}\bigg)^{2} - e^{-a p^{0}}~ \frac{M}{r^{2} (1 + \frac{M}{2 \, r})}  \bigg(\frac{d \, r}{d \, \tau}\bigg)^{2} \\ - e^{-a p^{0}}~\Bigg( r - \frac{M}{1 + \frac{M}{2 \, r}} \Bigg) \, \Bigg\{ \Bigg(\frac{d \, \theta}{d \tau} \Bigg)^{2}  + \sin^{2}\theta \Bigg(\frac{d \, \phi}{d \, \tau} \Bigg)^{2} \Bigg\} = 0. \label{geod3}
		\end{multline}
Similarly, we obtain the equations for the angles ($\theta$ and $\phi$)	
		\begin{equation}
		\frac{d^{2} \theta}{d \tau^{2}}  +   e^{-a p^{0}} \Bigg\{\frac{2}{r} \, \Bigg( \frac{1 - \frac{M}{2 \, r}}{1 + \frac{M}{2 \, r}} \Bigg)\frac{d \, r}{d \tau} \frac{d \, \theta}{d \tau} - \sin\theta \cos\theta \Bigg(\frac{d \, \phi}{d \tau} \Bigg)^{2}\Bigg\} = 0, \label{geod4}
		\end{equation}
and 		
		\begin{equation}
		\frac{d^{2} \phi}{d \tau^{2}} +e^{-a p^{0}}\Bigg\{\frac{2}{r} \Bigg( \frac{1 - \frac{M}{2 \, r}}{1 + \frac{M}{2 \, r}} \Bigg)  \frac{d \, r}{d \tau} \frac{d \, \phi}{d \tau} + 2 \, \cot\theta \frac{d \, \phi}{d \tau} \frac{d \, \theta}{d \tau}\Bigg\} = 0. \label{geod5}
		\end{equation}
In all these equations, we have the $a$ (deformation parameter) dependent terms due to the non-commutativity of the space-time. Here, we have considered correction terms for all orders of $a$. In the limit $a \rightarrow 0$, we get the commutative geodesic equations for the Schwarzchild black hole \cite{Diemer}. 

Since it is expected that the modifications due to the non-commutativity are very small, we take $ap^{0}\simeq .01$ (where $p^{0}$ is associated with the mass of a black hole). We numerically solve the above geodesic equations for different initial values of position, angular momentum and velocity. As is well known from previous work \cite{wang}, the orbits of the massive particles can be classified into different categories based on the initial values of position, angular momentum and velocity. There can be either bound orbits or unbound orbits. Amongst the unbound orbits, there are two possibilities, the escape orbits and the capture orbits. 
We compare orbits with the same initial values in the undeformed and deformed cases. We find, in general that the change due to the deformation parameter is uniform for various cases. As a representative case, we plot the trajectory of the massive test particle in a bound orbit around the black hole in Fig-1.

\begin{figure}[H]
\centering
 \includegraphics[scale=.45]{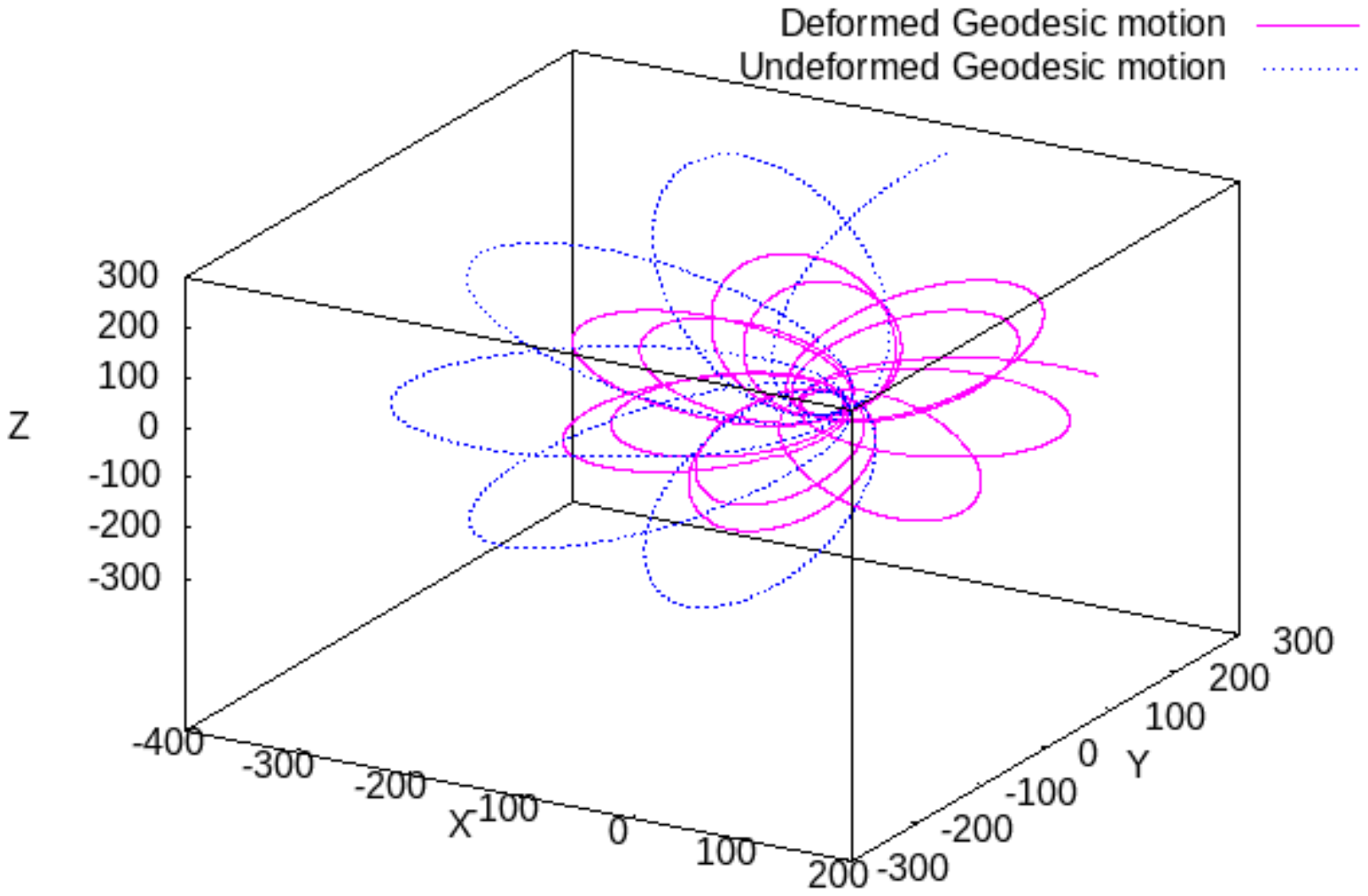}
\caption{Geodesic motion of a massive test particle around the black hole. The dotted line (blue color) represents the commutative space-time while the solid line (purple color) represents the non-commutative case.}
\label{fig1}
\end{figure}

Fig-1 shows the trajectories of a massive test particle around the black hole for both commutative (represented in blue color, dotted line) and non-commutative cases (represented in purple color, solid line). Here, the deviation in trajectory for the $\kappa$-deformed case compared to the commutative case is very distinct. The radius of the elliptical trajectory is reduced due to the non-commutativity of space-time. Thus, the particles will take less time to complete one elliptical path around the black hole in the $\kappa$-deformed space-time. 

The change in the precessing ellipse also means that the points of apogee and perigee marked by the sharp change in angular momentum are different in the $\kappa$-deformed Schwarzchild space-time. For bound orbits, the change in the angular momentum with time can be obtained by solving Eq.(\ref{geod2}) to Eq.(\ref{geod3}). We plot this evolution in Fig-2. 
\begin{figure}[H]
\centering
 \includegraphics[scale=.55]{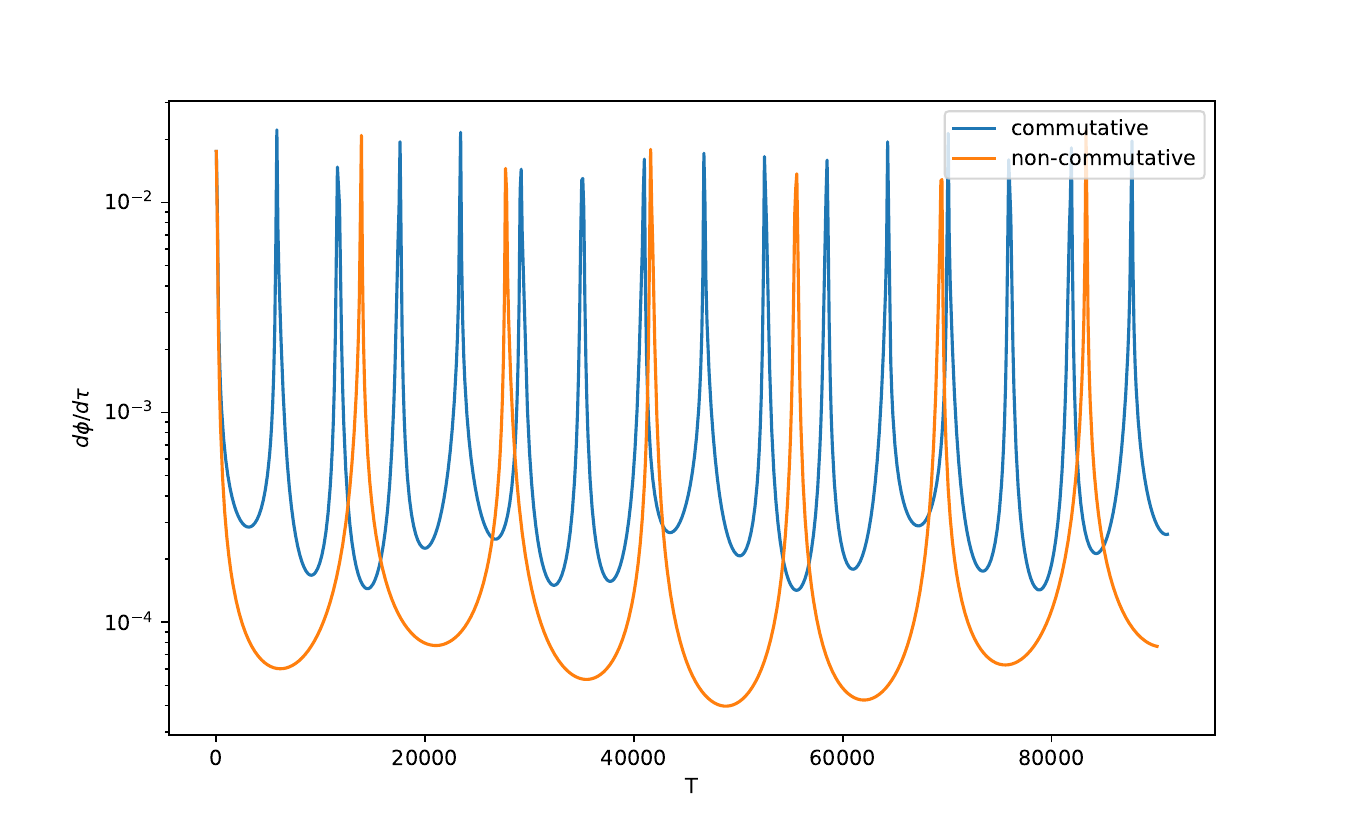}
\caption{The change in the angular velocity of the object around the black hole with time}
\label{fig2}
\end{figure}
As it is well known, apart from the energy, the angular momentum of the massive particle coming from infinity decides whether it will fall into the black hole or move in various trajectories about the black hole. It may also be scattered away from the black hole by the centrifugal barrier. So the angular momentum is quite important in determining the flow of the streaming particles around the black hole. We have plotted Fig-2 for the particle orbits shown in Fig-1. From Fig-1, we note that the particle moves around the black hole in a precessing elliptical trajectory. As we can see from the figure, the orbits in the deformed space-time have a lower angular momentum but the change in magnitude is not as significant as the change in the radius of the trajectory which is clearly visible from Fig-1. The major difference however is in the apsidal precession of the orbits. For a circular orbit, the plot of $d\phi \over d\tau$ should be a regular periodic oscillation of constant amplitude, however the apsidal precession of the orbits of particles around a Schwarzchild black hole leads to a change in the maximum and minimum of the angular momentum even in the case of the Scwharzchild black hole in commutative space-time. For the black hole in the deformed space-time, it is intriguing to note that the periodicity of the oscillations of the angular momenta changes significantly. Moreover, though the radius of the orbits are smaller, the particles tend to spend longer time in completing the orbits which means that they move slower around the black hole in the deformed space-time. This will mean that particles in bound orbits in $\kappa$ deformed space-time will move slower than the particles in an undeformed space-time and will therefore remain closer to the black hole. In the next subsection, we concentrate on the most important bound orbit of the particles, the innermost stable circular orbit ($r_{ISCO}$).

\subsection{Deformed effective potential and the ISCO}

The effective potential $V_{eff}$ is an important concept for the explanation of particle motion around a black hole because it defines the dynamics between centripetal force and gravitational force acting on the particle moving around the black hole. The $V_{eff}$ curve explains the nature of stability of orbital motion of particles around the black holes, which tells us whether the particle will spiral into the black hole or the particle will revolve around the black hole in a stable orbit. As we are looking at the geodesic nature of particles moving around the Schwarzchild's black hole in $\kappa$ deformed space-time, we calculate the effective potential for spherically symmetric geometry (taking $\theta = \pi/2 $). 

We start with the Lagrangian for a point particle in $\kappa-$deformed space-time, it is given by 
$\mathcal{\hat{L}} = \frac{1}{2} \hat{g}_{\mu \nu}(a) \dot{\hat{x}}^{\mu} \dot{\hat{x}}^{\nu} = - \frac{1}{2} \delta $, where $ \dot{\hat{x}}^{\mu} = \frac{d\hat{x}^{\mu}}{d\sigma}$. Here, $\delta = 0, 1$ is for massless and massive particles, respectively. We obtain the Lagrangian $ \hat{\mathcal{L}}$ as
\begin{equation}
 \mathcal{\hat{L}} = \frac{1}{2} \bigg(- f(r)\dot{t}^{2} \, + \frac{l(r)}{f(r)} \, \, e^{-4 a p^{0}} \, \dot{r}^{2} \, + \, \frac{l(r)}{f(r)} \, \, e^{-4 a p^{0}} \, r^{2} \, \dot{\theta}^{2} \, + \, \frac{l(r)}{f(r)} \, \, e^{-4 a p^{0}} \, r^{2} \sin^{2}\theta \, \dot{\phi}^{2} \bigg)  = - \frac{1}{2} \delta .
\end{equation}
Using expressions of $f(r)$ and $ l(r)$ and rewrite the above equation as
\begin{multline}
\mathcal{\hat{L}} = \frac{1}{2} \Bigg\{ - \Bigg( \frac{1 - \frac{M}{2 r}}{1 + \frac{M}{2 r}}\Bigg)^{2} \dot{t}^{2} + \bigg( 1 + \frac{M}{2 r}\bigg)^{4} e^{-4 a p^{0}}  \dot{r}^{2}  \\
+  \bigg( 1 + \frac{M}{2 r} \bigg)^{4} e^{-4 a p^{0}}  r^{2} \dot{\theta}^{2} + \bigg( 1 + \frac{M}{2 r}\bigg)^{4} e^{-4 a p^{0}} r^{2} \sin^{2}\theta \dot{\phi}^{2} \Bigg\} = - \frac{1}{2} \delta . \label{L_particle}
\end{multline}
Using the above equation, we obtain the constant equation of motion for $ \mu = t, \phi $ as
\begin{equation}
	\Bigg( \frac{1 - \frac{M}{2 r}}{1 + \frac{M}{2 r}} \Bigg)^{2} \, \dot{t} = k,
\end{equation}
and
\begin{equation}
	\Bigg(1 + \frac{M}{2 r}\Bigg)^{4} \, r^{2} \, e^{-4 a p^0} \, \dot{\phi} = \hat{h}.
\end{equation}
Here, $k$ and $\hat{h}$ are constants. We are confining to a fixed plane by choosing  $\theta = \pi/2 $, the equatorial plane, thus the effective potential will only give us knowledge about the trajectories in the equatorial plane. From the conserved quantities energy, $E$ and total angular momentum $\hat{L}$, we obtain 
\begin{equation}
E = \, f \, \frac{d t}{d \tau} = k \, \, \, ; \, \, \, \hat{L} = \frac{l}{f} \, r^{2} \, e^{-4 a p^{0}} \, \frac{d \phi}{d \tau} = \hat{h} .
\end{equation}
Substituting these constants of motion in Eq.(\ref{L_particle}), we find
\begin{equation}
\bigg(\frac{d r}{d \tau}\bigg)^{2} =  e^{4 a p^{0}}\frac{1}{l} \bigg( E^{2} - V_{eff}^{NC}(r) \bigg),
\end{equation}
where, the non-commutative ($\kappa$-deformed) effective potential is
\begin{equation}
	V_{eff}^{NC} (r) = f(r) \Bigg( \delta + e^{4 a p^0}\frac{f}{l}\frac{\hat{L}^{2}}{ r^{2}} \Bigg) .
\end{equation}
Thus, we obtain the effective potential for the non-commutative space-time geometry($\kappa$-deformed) as
\begin{equation}
	V_{eff}^{NC} (r) = \Bigg( \frac{1 - \frac{M}{2 r}}{1 + \frac{M}{2 r}} \Bigg)^{2} \Bigg( \delta + e^{4 a p^{0}} \frac{\hat{h}^{2}}{ r^{2} \, (1 + \frac{M}{2 r})^{4}} \Bigg) . \label{V_{eff}}
\end{equation}
From the above equation, we obtain $V_{eff}^{NC}$ for massless particle ($\delta = 0$) as
\begin{equation}
	V_{eff}^{NC} (r) = e^{4 a p^{0}}\Bigg( \frac{1 - \frac{M}{2 r}}{1 + \frac{M}{2 r}} \Bigg)^{2} \Bigg( \frac{\hat{h}^{2}}{ r^{2} \, (1 + \frac{M}{2 r})^{4}} \Bigg) . \label{del0}
\end{equation}
Similarly, for massive particle ($\delta = 1$), we find effective potential as
\begin{equation}
	V_{eff}^{NC} (r) = \Bigg( \frac{1 - \frac{M}{2 r}}{1 + \frac{M}{2 r}} \Bigg)^{2} \Bigg( 1 + e^{4 a p^{0}}\frac{ \hat{h}^{2}}{ r^{2} \, (1 + \frac{M}{2 r})^{4}} \Bigg) . \label{del1}
\end{equation}
In the limit, $a \rightarrow 0$, we get the commutative result \cite{Diemer}.
By plotting Eq.(\ref{del0}) and Eq.(\ref{del1}), we obtain the curve for the effective potential versus the logarithmic scale of radius from the center of the point black hole. We obtain figures 3(a) and 3(b) for massless and massive particles, respectively. For both massive and massless particles, we find that the effective potential curves for lower values of angular momentum $\hat{L}$, the difference in deformed and undeformed space-time geometry is negligible. As the angular momentum increases, the difference in the effective potential curve for space-time geometry with different angular momentum also increases. In the massless case, as $r$ increases, the effective potential increases reaches a maximum and then decreases to the minimum of the potential, while for the massive particles,  as $r$ increases, the effective potential decreases and reaches a minimum and then again rises and keeps going  to infinity.
\begin{figure}[H]
	\centering
	\begin{subfigure}{0.49\linewidth}
		\includegraphics[width=\linewidth]{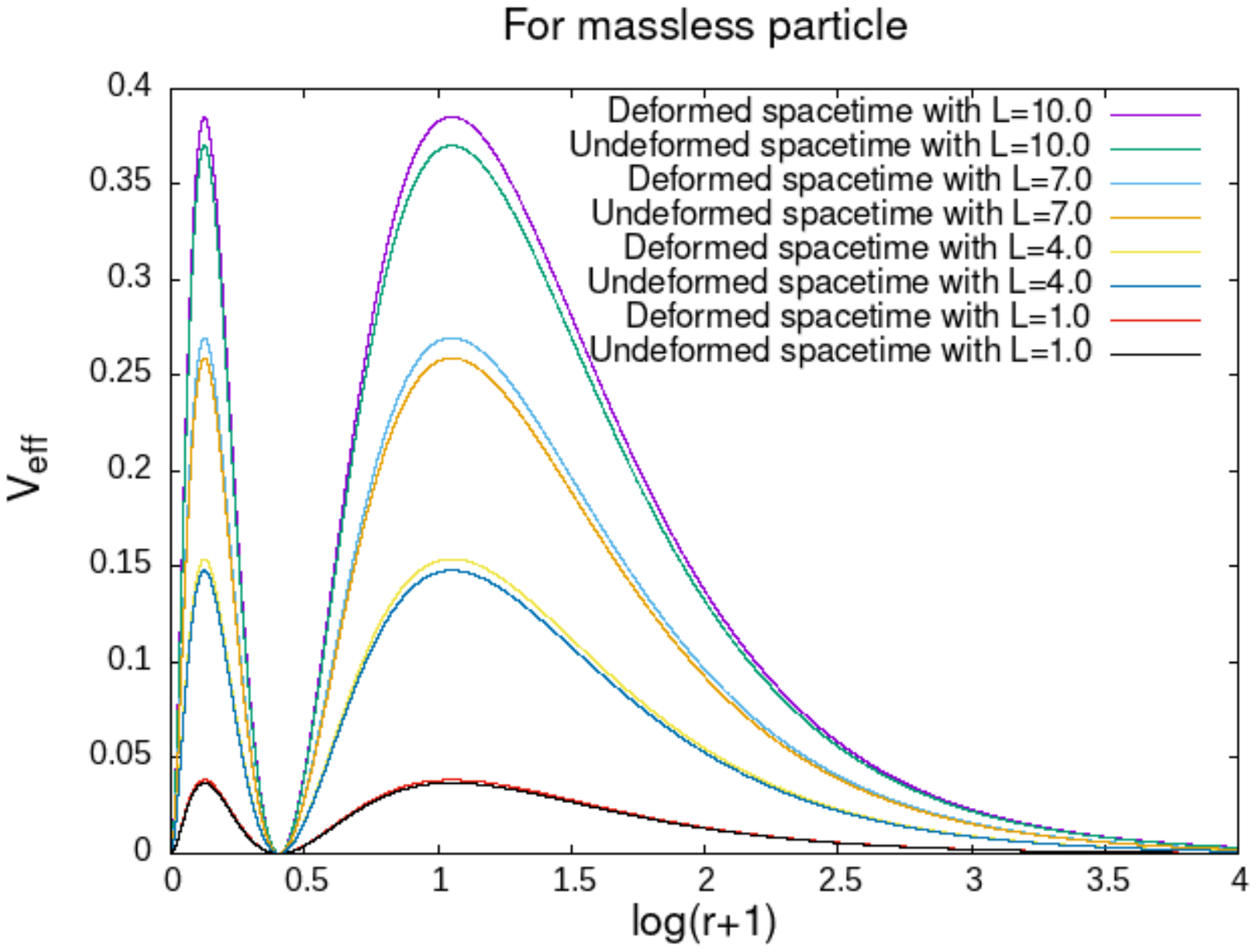}
		\caption{For massless particle}
	\end{subfigure}
	\begin{subfigure}{0.49\linewidth}
		\includegraphics[width=\linewidth]{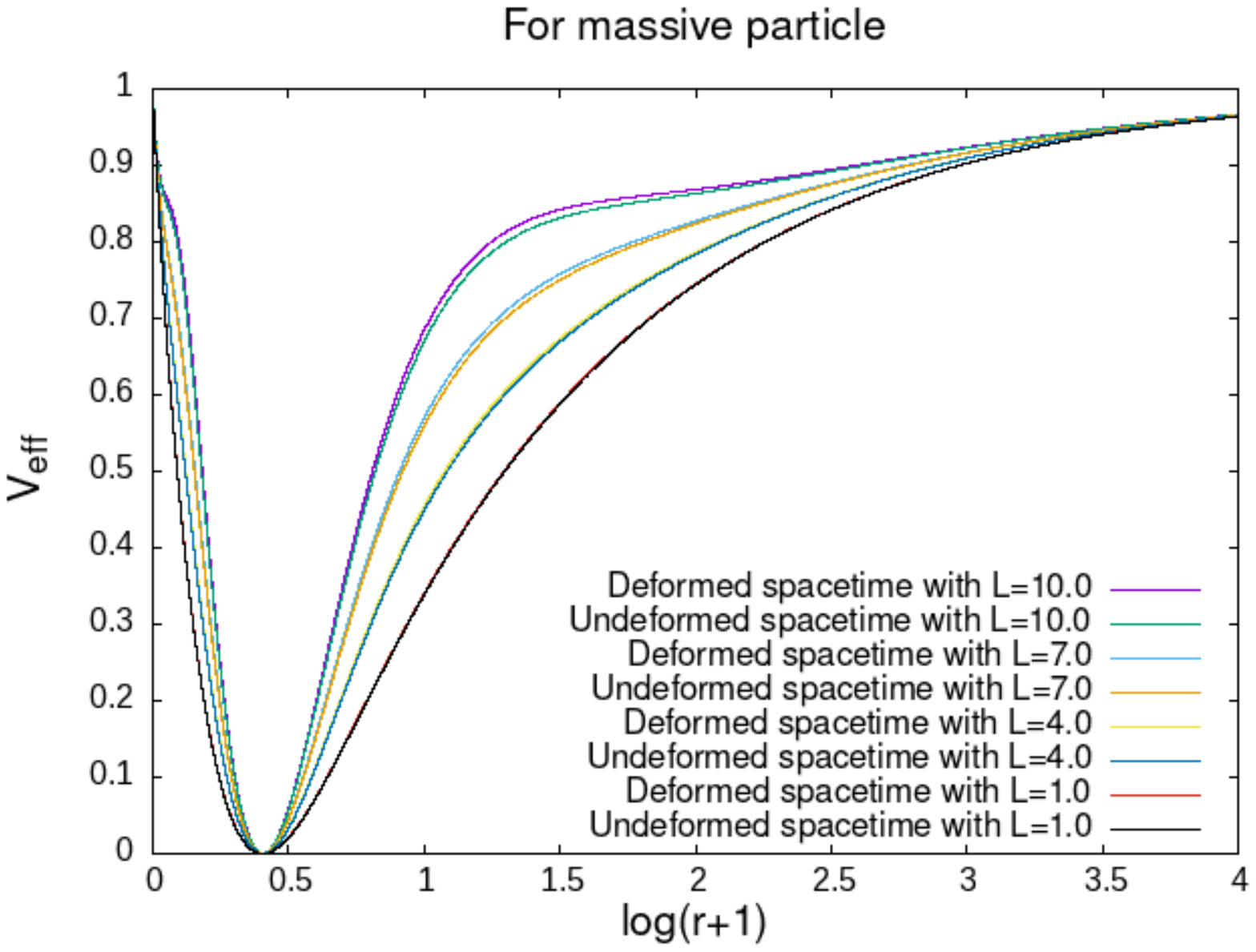}
		\caption{For massive particle}
	\end{subfigure}
	\caption{The graph shows the profile of the effective potential $V_{eff}$ versus $log(r+1)$ in the undeformed and deformed space-time of the black hole. The plots are for several values of the angular momentum; (a) for massless and (b) for massive particle.}
	\label{fig:$V_{eff}$ versus $log(r+1)$}
\end{figure}
From this detailed study we can conclude that the massive particle orbits will be affected more by the $\kappa$ deformed space-time. 

The dynamics of moving particles around black holes is crucial, and for understanding this behavior of such systems, the analysis of the Innermost Stable Circular Orbit is required. The radius of the innermost stable circular orbit $r_{ISCO}$ defines the smallest radius of the stable circular orbit where particles revolve around the black hole due to the presence of gravitational force.

Since we are looking at a circular orbit, we take the derivative of the effective potential Eq.(\ref{V_{eff}}) with respect to $r$ and set it to zero ($\frac{d V_{eff}^{NC}}{d r}=0$), as
\begin{multline}
	\delta \frac{(1 - \frac{M}{2 r})(\frac{M}{r^{2}})}{(1 + \frac{M}{2 r})^{2}} +  \delta \frac{(1 - \frac{M}{2 r})^{2}(\frac{M}{r^{2}})}{(1 + \frac{M}{2 r})^{3}}  +  e^{4 a p^{0}}\frac{\hat{h}^{2}}{r^{2}} \frac{(1 - \frac{M}{2 r})(\frac{M}{r^{2}})}{(1 + \frac{M}{2 r})^{6} }\\   + e^{4 a p^{0}}\frac{\hat{h}^{2}}{r^{2}} \frac{3 (1 - \frac{M}{2 r})^{2}(\frac{M}{r^{2}})}{(1 + \frac{M}{2 r})^{7}} -  e^{4 a p^{0}}\frac{2 \hat{h}^{2}}{r^{3}} \frac{(1 - \frac{M}{2 r})^{2}}{(1 + \frac{M}{2 r})^{6}} = 0
\end{multline}
We rewrite the above equation as
\begin{equation}
	2 \delta M r^{2} - e^{4 a p^{0}} 2 \hat{h}^{2} \frac{(1 - \frac{M}{2 r})}{(1 + \frac{M}{2 r})^{3}} r + e^{4 a p^{0}}\frac{2 \hat{h}^{2} M (2 - \frac{M}{2 r})}{(1 + \frac{M}{2 r})^{4}} = 0 .
\end{equation}
Considering, $\frac{M}{2 r} << 1$, we find
\begin{equation}
	(\delta M) r^{2} - (e^{4 a p^{0}} \hat{h}^{2}) r + (e^{4 a p^{0}} 2 \hat{h}^{2} M) = 0 .
\end{equation}
Solving for the roots of a quadratic equation of $r$, we obtain
\begin{equation}
	r = \frac{\hat{h} e^{2 a p^{0}}}{2 \delta M} \bigg\{ \hat{h} e^{2 a p^{0}} \pm \sqrt{\hat{h}^{2} e^{4 a p^{0}} - 8 \delta M^{2} } \bigg\} . \label{roots}
\end{equation}
By setting  $\hat{h}^{2} e^{4 a p^{0}} - 8 \delta M^{2} = 0$, ensures that the effective potential has a local minimum at the smallest possible radius, indicating stability, which is a necessary condition for the ISCO\cite{Hobson}, we get
\begin{equation}
	\hat{h}^{2} = 8 \delta M^{2} e^{-4 a p^{0}} .
\end{equation}
Using the above value in Eq.(\ref{roots}) gives
\begin{equation}
	r_{ISCO} = 4 M .
\end{equation}
Note that here, the radius of the innermost stable circular orbit ($r_{ISCO}$) of an object does not get modified due to the presence of non-commutativity of the space-time.
The $r_{ISCO}$ for the non-rotating black hole, such as a Schwarzchild black hole, is obtained approximately $\frac{6 G M}{c^{2}}$, where $G$ is the gravitational constant, $M$ is the mass of the black hole, and $c$ is the speed of light \cite{Hobson}. In our case for $\kappa-$deformed space-time geometry, we obtain  $r_{ISCO} = 4 M $, considering $G = c = 1$ in astrophysical units.

\subsection{Particle concentration around the black hole}

So far in our study of the geodesics, we have found that though the effective potential does not change significantly in the $\kappa$ deformed space-time, the individual particle orbits are strongly affected by the 
deformity in the space-time. To study whether the change in the trajectories affect particles streaming around black holes, we do a multi-particle simulation of a large number of particles moving in the  $\kappa$ deformed space-time around the black hole. 
We consider freely streaming particles coming from a distant point and moving towards the black hole. Since we expect more differences for higher angular momentum values, we choose particles whose initial angular momentum is high.  

For the simulation, we generate a large number of particle randomly around the black hole. Since we have to define a physical space, we use the value of the ISCO that we have obtained in the previous section. We put the particles randomly within a large circle of radius larger than the ISCO. After giving an initial position and initial velocity, we numerically obtain the positions of the particles at later times. We find that the particles travel along their geodesics and move around in definite trajectories based on the initial value of their angular momentum and energy. Individual snapshots at different times will give us different patterns of the particles. Moreover, since the initial radius, angular velocity are all randomly generated it is not necessary that all the particles will move in bound trajectories. We had previously observed that changing the initial values often results in getting open and unbound trajectories of the particles. So a majority of the particles in the simulation will have unbound trajectories. Only a few of them will be having a bound trajectory, hence within a fixed area, the particle number is not conserved. 

In our simulations, we have taken about $10^{6}$ particles. To study the distribution of these particles in the deformed and undeformed space time, we need to count the number of particles in a particular area at different time intervals. Since, the space around the Schwarzchild black hole is usually characterized in terms of the $(r, \theta, \phi)$ coordinates, we choose to do the counting by generating annular rings of bins around the black hole. This annular bins are sketched in the plane of the paper in Fig-4. for better clarity.  
\begin{figure}[H]
\centering
 \includegraphics[scale=.25]{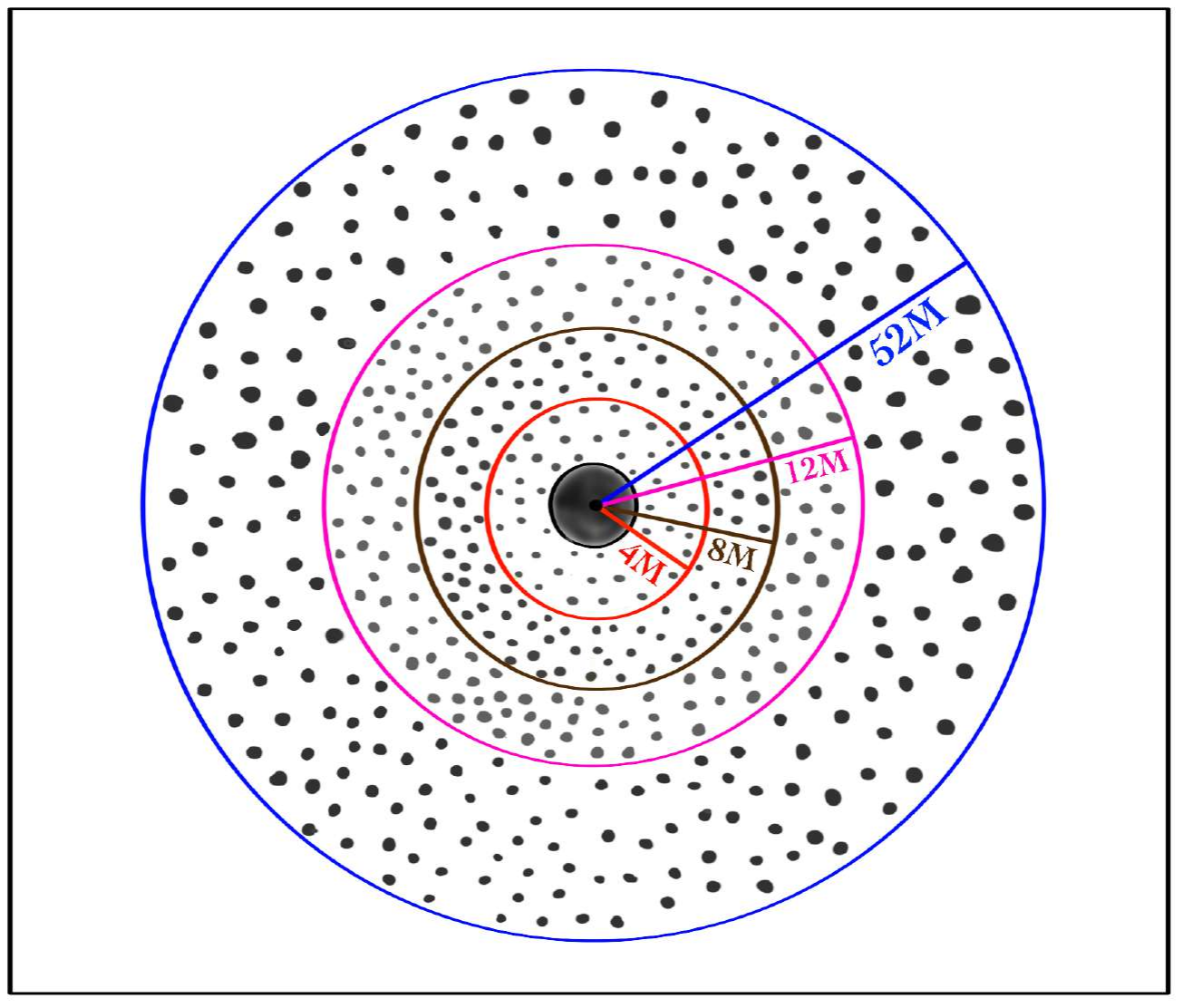}
\caption{Sketch of the annular rings around the black hole required for binning the data. The innermost radius is equal to the innermost stable circular orbit, and the radii of each ring is then increased by 4 units. Not all the rings are shown in the figure. The first three rings are labelled by their radii $4M$, $8M$, $12 M$ and the largest ring is labelled at $52 M$. The sketch is an illustration of our binning method and is not drawn to scale. }
\label{fig4}
\end{figure}
Each of the annular rings is labelled from $1$ to $14$, in increasing values of the radius, the spherical polar coordinates are used to calculate the radius. Unfortunately, in this method the angular information about the distribution are not manifested. However, we still find some interesting results. 
 
We start with an initial random distribution of particles in all the annular rings. The initial distribution is shown in Fig-5. For both the deformed and the undeformed space-time we have the same distribution.
\begin{figure}[H]
\centering
 \includegraphics[scale=.35]{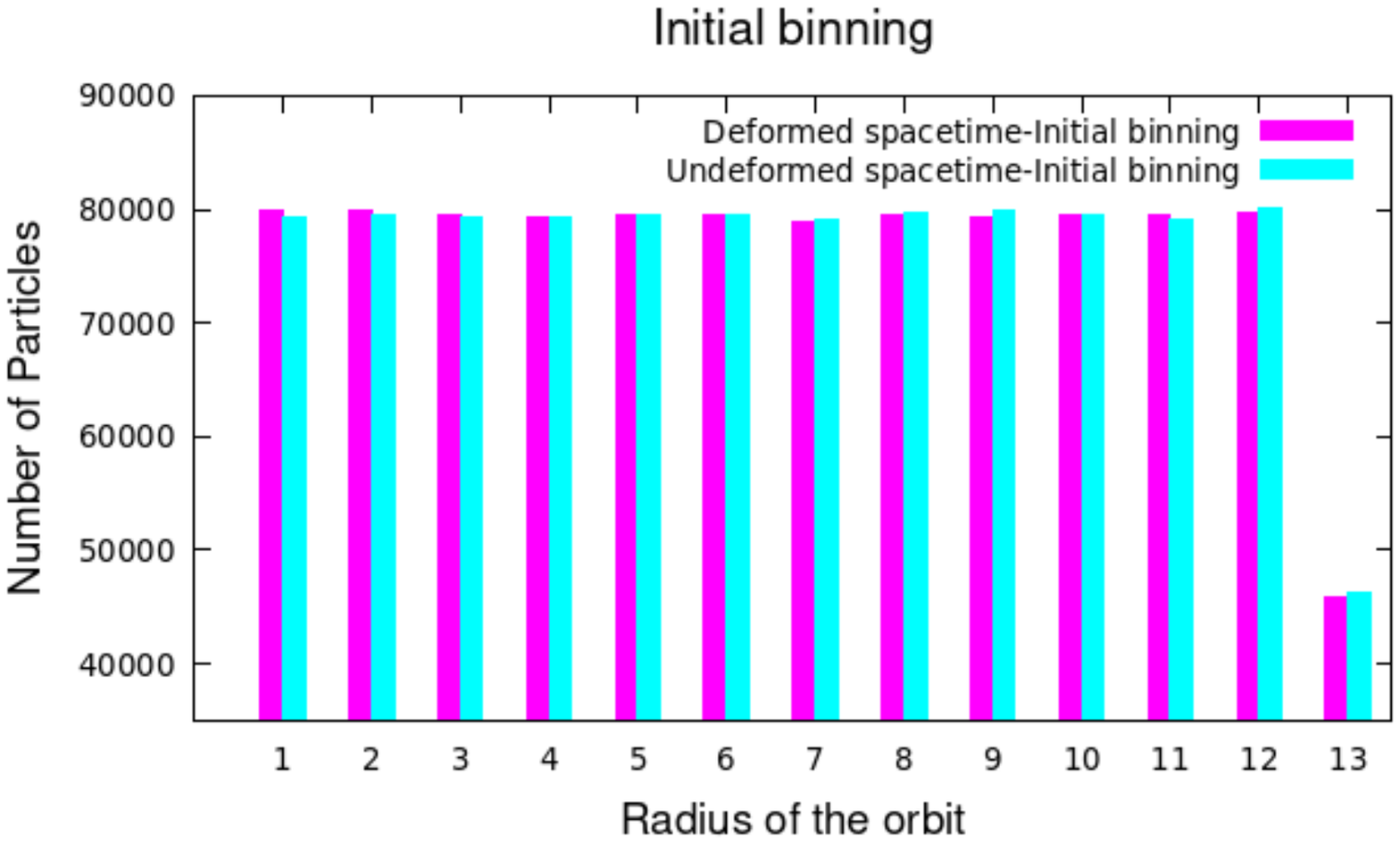}
\caption{Particles are randomly distributed at $t = 0$ around the black hole. When we bin the particles in annular bins, we find that the number of particles in each bin is approximately equal. This is done for both the undeformed as well as the deformed space-time.}
\label{fig5}
\end{figure}

We see from the above bar diagram, the distribution shows approximately equal number of particles in all the bins. The x-axis has the bin distribution. The bin corresponding to number $1$ has particles close to the ISCO (around $4$ M). Each bin has an increasing value of radius. The average radius of the particle in the bin can be obtained by multiplying the bin number by $4$. So the radius of the particles which are farthest from the black hole is approximately $52$ M. The initial position and velocities of the particles are generated by using a standard random number generator. The only constraint that we have put is that the initial angular momentum values are approximately close (but not exactly the same) as those for which we have already plotted geodesic curves similar to the one shown in Fig-1. We then evolve the system for a certain length of time. The final results are shown in Fig-6.   
\begin{figure}[H]
	\centering
	\begin{subfigure}{0.47\linewidth}
		\includegraphics[width=\linewidth]{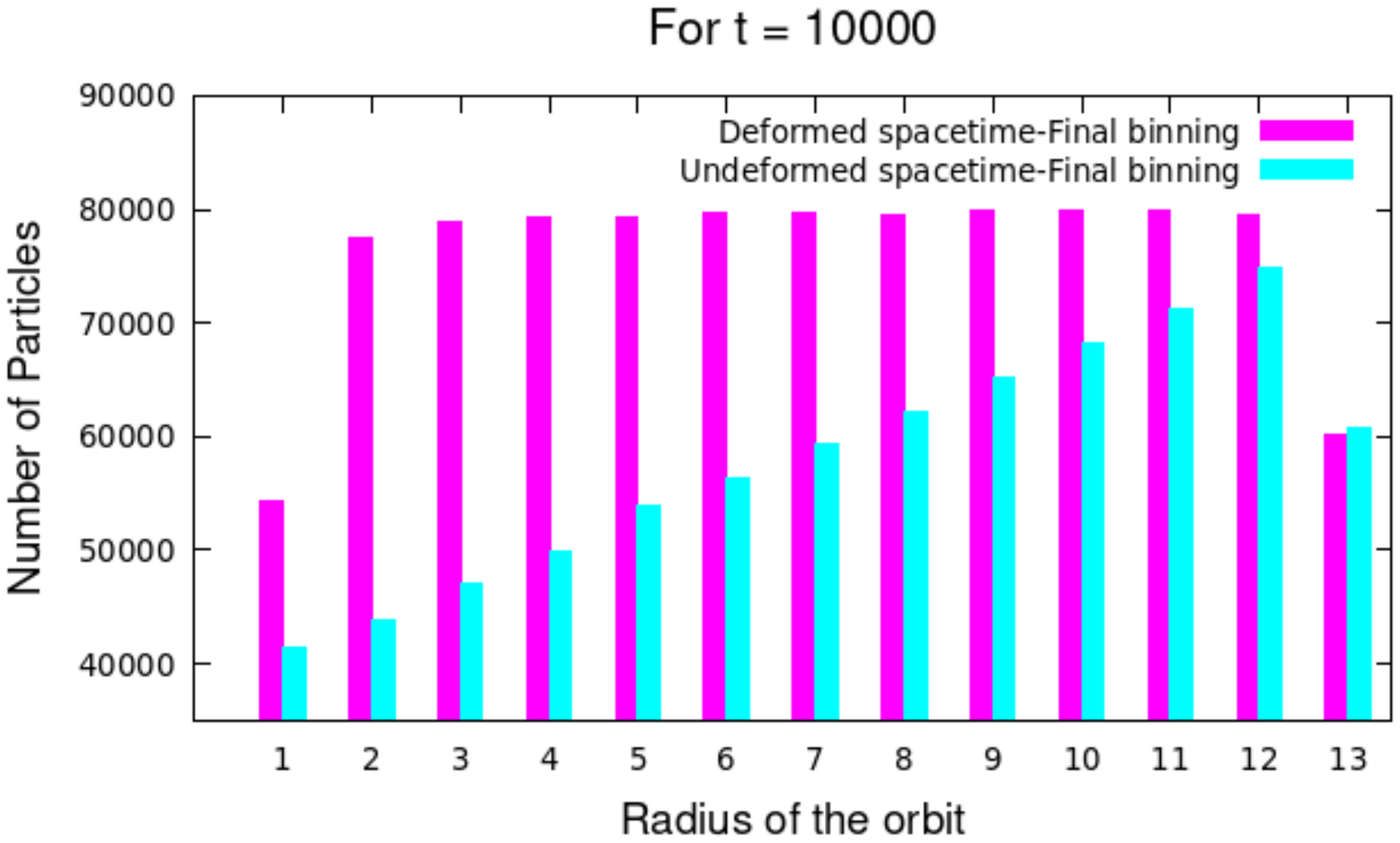}
		\caption{Snapshot at $t = 10^4$}
	\end{subfigure}
	\begin{subfigure}{0.47\linewidth}
		\includegraphics[width=\linewidth]{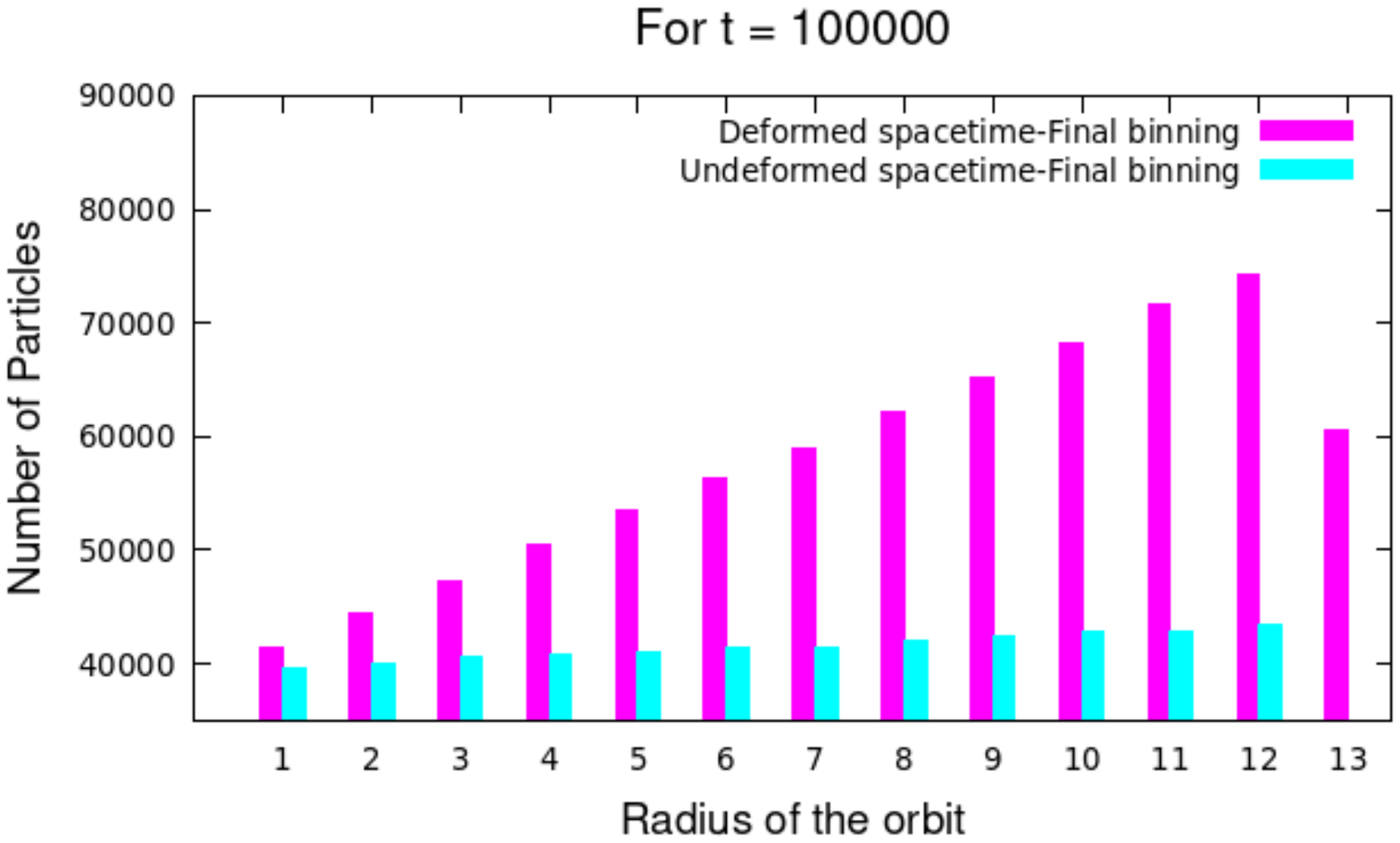}
		\caption{Snapshot at $t = 10^5$}
	\end{subfigure}
	\caption{The graph illustrates the profile of final particle accumulation in the surrounding of the black hole in the presence of undeformed and deformed space-time geometry. For the deformed space-time, the particle distribution is distinctly different compared to the undeformed space time }
	\label{fig:Binning}
\end{figure}

The two figures show the result of the particle streaming  at two different times. On the left, Fig-6.a shows the results after a total number of timesteps $t = 10^4$, and on the right, Fig-6.b, shows the distribution at time $t =10^5 $ for both the deformed and the undeformed spaces. Overall we see that there is a large number of particles in unbound orbits which move out of the area that we are considering for counting the particles. The initial count of particles show approximately $80000$ particles distributed in the bins around the ISCO.   
 
In the undeformed commutative space-time, the particles tend to move away from the black hole, so the particle concentration in the lower numbered bins seem to decrease. This means that most of the particles are in unbound orbits. For the same intial conditions, in the deformed case, the particles tend to remain closer to the black hole. Though they also move away from the black hole eventually, they are slowed down by the deformation of the space-time. All these particles are in the unbound trajectories. There are some particles in the bound orbits and we see that even in large time, the initial bins have about $40000$ particles. This happens both for the deformed and the undeformed space time. The streaming particles beyond the ISCO are thus affected by the deformation of the space time around the black hole. 
This indicates that the accretion of particles in $\kappa$ deformed space-time would vary considerably based on the deformation parameter. 
Our results are similar to the results obtained for accretion onto a non-commutative-inspired Schwarzschild black hole \cite{sunandan} where the authors found that the matter accretion rate increases in non-commutative space-time. In our case too, we have found that massive particles tend to move slower in the non-commutative background. For the same values of constants of motion, in the deformed space-time we have a higher density of particles closer to the black hole. A more detailed study may lead to more interesting results on accretion of particles in $\kappa$ deformed space-time around Schwarzchild black holes.  

\section{Conclusions}

In this work, we have studied the behavior of a particle moving very close to a static spherical black hole in $\kappa$ deformed space-time. For this, we have constructed the $\kappa$-deformed metric for the Schwarzchild black hole. We find the geodesic equations using this $\kappa$-deformed metric. We analyze these geodesic equations to determine the path of a test particle around the Schwarzchild black hole. We investigate how the particle's motion varies in the deformed and conventional scenarios. The modified Lagrangian is then obtained using the $\kappa$-deformed metric, and from this, we derive the $\kappa$-deformed effective potential, valid up to the first order in $a$. Using this, we obtain varied trajectories based on the angular momentum values. The radius of the innermost stable circular orbit ($r_{ISCO}$) is calculated using a modified effective potential. The particle distributions surrounding the Schwarzchild black hole in $\kappa$-deformed space-time are then analyzed using an annular binning technique.

Our investigation into the dynamics of particle motion and distribution around a Schwarzchild black hole within the framework of $\kappa$-deformed space-time geometry has provided valuable insights into the interplay between gravity and quantum effects. Our study of geodesic motion shows the deviation in trajectories for an object around the $\kappa$-deformed Schwarzchild black hole. The reduction in the radius of trajectories and the enhancement of the period for completing orbits highlight  
the effect of non-commutativity on the streaming of free particles. 
We have also obtained the $\kappa$-deformed effective potential. This modified effective potential demonstrates the influence of the $\kappa$-deformation parameter on the stability of particle orbits around the black hole. Differences in effective potential curves between $\kappa$-deformed and undeformed space-time geometries indicate that the presence of the non-commutative parameter affects the orbits of the particles. 

We have found the innermost stable circular orbit (ISCO) using the effective potential expression. This critical parameter defines the smallest stable orbit radius, where particles revolve around black holes due to gravitational forces. We found that for the non-rotating Schwarzchild black hole, the $r_{ISCO}$ is $4$M, this did not change due to the non-commutativity of the space-time. A similar study has been done in \cite{Cruz,mandal}, in ref. \cite{Cruz}, the authors have studied the geodesic structure of AdS Schwarzchild's black hole and analyzed the effective potential to determine the possible motions of massive particles. In \cite{mandal}, the geodesic motion of the particles has been studied for improved Schwarzchild black hole. The authors analyzed the effective potential to study the trajectories of the particles, and also found the modified $r_{ISCO}$. Finally, we analyzed the distribution of a large number of particles in the vicinity of the $\kappa$-deformed black hole. We found that the particles tend to stay closer to the black hole in the non-commutative space-time. This is because though the particles move in trajectories with smaller radii, they also tend to move slower in the non-commutative space-time. Due to this slowing down of the particles, the particles remain closer to the black hole for a longer time.  

The accretion of matter around black holes releases a vast quantity of energy in the form of radiation. The release of energy during the accretion process may affect star formation rates and the overall dynamics of the galaxy. A slow accretion rate means less material falling into the black hole, resulting in a lowering of the luminosity for the outward radiation \cite{Liu, Hong-Yang}. Slow accretion rates can impact the emission of gravitational waves from the black hole system \cite{Font}. Gravitational waves are primarily generated by events like black hole mergers and neutron star collisions. The accretion process can also produce gravitational waves. Observing these waves can help us understand the black hole's characteristics and accretion dynamics. The Hawking radiation theory states that black holes release radiation as a result of quantum phenomena near the event horizon. A slow accretion rate may affect the equilibrium between accretion and Hawking radiation, influencing the black hole's overall mass and evolution. In our investigation, we found that the existence of non-commutativity causes particles to move slowly. Thus, we can infer that the slow accretion process caused by non-commutativity has the same implications as discussed above.

Overall, this work elucidates the effects of space time non-commutativity on particle motion and their distribution in $\kappa$-deformed space-time around a Schwarzchild black hole. Our study contributes to the ongoing exploration of the fundamental principles governing the dynamics of objects in the vicinity of black holes. We plan to extend this work by studying and understanding the accretion of particles in non-commutative space time, more specifically to the case of the  $\kappa$-deformed space-time which may lead to further interesting results. 

\begin{center}
 Acknowledgments
\end{center}  
The authors would like to thank  E. Harikumar for useful discussions and comments.
The computational part of this research work was carried out on the computational facility  set up from funds given by the DST- SERB Power Grant no. SPG/2021/002228 of the Government of India. D.K acknowledges financial support from DST- SERB Power Grant no. SPG/2021/002228 of the Government of India. S.K.P thanks UGC, India, for the support through the JRF scheme (id.191620059604).   During the initial part of this work A.S was supported by the University of Hyderabad IoE Grant No UoH-IoE-RC5-22-020. A.S. is currently supported by the National Natural Science Foundation of China with grants Nos.12247107, 12075007.

\section*{Data availability}

Data sharing is not applicable to this article as no data sets were generated or analyzed during the current study.

\section*{Conflict of interest} 

The authors declare no conflict of interest.

\end{document}